\documentstyle[prl,aps,epsfig,twocolumn]{revtex}  

\begin{document}

\draft

\title{Microoptical Realization of Arrays of Selectively Addressable Dipole Traps:\\
A Scalable Configuration for Quantum Computation with Atomic Qubits}

\author{R. Dumke, M. Volk, T. M\"uther, F.B.J. Buchkremer, G. Birkl, and W. Ertmer}

\address{Institut f\"ur Quantenoptik, Universit\"at Hannover,
Welfengarten 1, D-30167 Hannover, Germany}

\date{\today}
\maketitle

\begin{abstract} 

We experimentally demonstrate novel structures for the realisation of registers 
of atomic qubits: 
We trap neutral atoms in one and two-dimensional arrays
of far-detuned dipole traps
obtained by focusing a red-detuned laser beam with a microfabricated
array of microlenses.
We are able to selectively 
address individual trap sites due to their large lateral separation of 125 $\mu$m. 
We initialize and read out different internal states for the 
individual sites. 
We also create two interleaved sets of  
trap arrays with adjustable separation, as required for many proposed implementations
of quantum gate operations.
%

\end{abstract}


\narrowtext


The dramatic progress in micro- and nanofabrication of the hardware for information
technology will lead to the encoding of logical information on a single
particle basis in the near future.
Therefore, the quantum behavior of the physical carriers of information has to be considered.
Moreover, quantum effects may provide qualitatively new modes 
for information processing driving research in the field of quantum information
processing \cite{STEA98}.



Among the broad range of currently investigated approaches, 
important progress has been obtained with atom physical schemes.
Entanglement and quantum gate operations have been achieved with trapped ions 
\cite{MONR95} or photons in
cavity QED experiments \cite{HAGL97}. 
Several schemes for 
quantum gates based on the
direct interaction of neutral atoms 
have been proposed theoretically \cite{JAKS99,BREN99,JAKS00,CHAR02}.
Still, there is an intensive search for appropriate systems allowing the experimental 
realisation of these schemes.


Several requirements have to be fulfilled for the sucessful implementation of quantum computation 
\cite{DIVI00}, such as 
the scalability of the physical system, the capability to initialize and read out qubits,
long decoherence times,  
and the existence of a universal set of quantum gates.  
These requirements are potentially met by setups which 
are based on miniaturized structures for trapping, guiding, and manipulating neutral atoms
as it is investigated in the newly developing field                                             
of integrated atom optics or {\it ATOMICS} \cite{ATOMICS}. 
This approach draws its strength from the combination of the well developed techniques
for the manipulation of atomic quantum states with the state-of-the-art manufacturing 
basis of micro- and nanofabrication.
A number of groups use microfabricated charged and current carrying 
structures for this purpose
\cite{WEIN,REIC99,MUEL99,DEKK00,KEY00,FOLM00,ZIMBEC,ENG}. 

As an alternative approach, 
we have proposed the application of microfabricated {\it optical} 
elements (microoptical elements) and have developed a number of configurations for 
atom optics, atom interferometry, and quantum 
information processing \cite{BIRK01,MOSKAU01}. We consider this approach to
be extremely powerful since most of the techniques for manipulating atomic qubits are
based on the interaction with light and since optical trapping potentials
can be made state selective in a simple fashion as required for many  
quantum gate proposals \cite{JAKS99,BREN99,JAKS00}.

\begin{figure}
   \begin{center}
   \parbox{7.5cm}{
   \epsfxsize 7.5cm
   \epsfbox{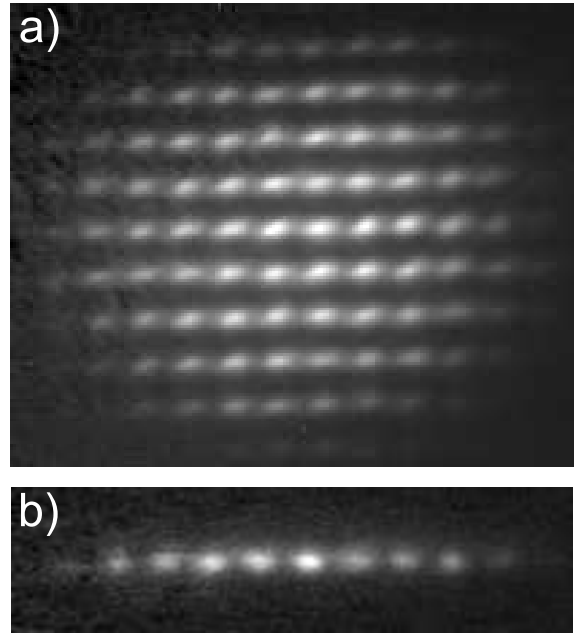}
}
   \end{center}
   
   \caption{(a) Two- and (b) one-dimensional arrays of rubidium atoms trapped in 
   arrays 
   of dipole traps. The traps are created using a microoptical lens array and 
   are separated by 
   125 $\mu$m.
   The brightest traps contain about $10^{3}$ atoms. }    
    \label{ARRAYFIG1}
\end{figure}

In this paper we present the first experimental implementation of 
microfabricated optical systems for quantum computing purposes with atoms:
Figure \ref{ARRAYFIG1} shows rubidium atoms trapped in one and          
two-dimensional arrays of 
dipole traps created by a microoptical system.
More than 80 traps hold atoms in Fig. 
\ref{ARRAYFIG1} (a).
Each trap can act as a memory site for quantum 
information encoded in the two hyperfine groundstates of the atoms. 
Thus, the arrays can serve as registers of atomic qubits. 

%
%
%

\begin{figure}
   \begin{center}
   \parbox{6cm}{
   \epsfxsize 6cm
   \epsfbox{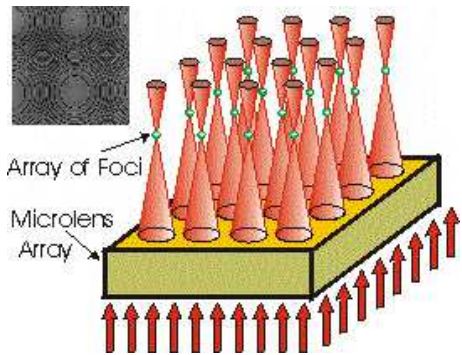}
}
   \end{center}
   
   \caption{A two-dimensional array of laser foci is 
   created by focusing a single laser beam with an array of microlenses. Inset: Phase 
   contrast image and typical cross section 
   of a small part of a diffractive microlens array.}        
    \label{ARRAYFIG2}
\end{figure}

We obtain the trap arrays by employing a 
two-dimensional array of spherical, diffractive microlenses with a  
focal length of 625 $\mu$m and a lens diameter and 
separation of 125 $\mu$m (inset in Fig. \ref{ARRAYFIG2}). 
The microlens array is made of fused silica and 
contains 50x50 diffractive lenslets. 
%
The trapping light is derived from a 500 mW amplified diode laser system and
is sent through a rubidium gas cell heated to a temperature of 
110$^{\circ}C$ serving as a narrowband absorption filter. 
This reduces the strong background of amplified spontaneous emission 
by at least two orders of magnitude otherwise preventing the 
operation of a dipole trap due to scattering of resonant photons. The light is 
then sent through an acoustooptical switching device and 
through a polarizer, which ensures a high degree of linear polarisation. 
The remaining light (typical power P = 100 - 200 mW, typical detuning $\Delta\lambda$ 
= 0.2 to 2 nm below the 
$5S_{1/2} (F=3) \rightarrow 5P_{3/2} (F'=4)$ transition at 780 nm ('red-detuning')) 
is focused by the microlens array. 
In order to have full 
optical access for atom preparation and detection, 
we image the focal plane of the microlens array onto a magnetooptical trap (MOT) 
with the help of two achromats 
(magnification = 1 and no significant reduction of numerical aperture).
Thus, we obtain an array 
of foci with a separation of 125 $\mu$m and a spot size of ($7 \pm 2) \mu$m ($1/e^2$-radius of intensity). 
For a red-detuned laser beam this results in an array of dipole traps, 
each analogous to a 
trap obtained by a single focused laser beam 
\cite{GRIM00,SCHL01} (see also \cite{BOIR98}).
The optical transfer of the trapping light has the additional advantage 
that we can place the microoptical system outside the vacuum chamber 
and thus can switch between
a variety of microoptical elements easily.
On the other hand, trapping of atoms directly in the first focal plane 
close to the surface is possible as well.

We load the array of dipole traps and detect the trapped 
atoms similar to \cite{KUPP00}:
We start with a MOT of $10^7$ to $10^8$ $^{85}Rb$ atoms 
which we overlap for several hundreds of ms with the dipole trap array and 
optimize the 
loading process for highest atom number.
The MOT is
then switched off and the atoms are held in the dipole traps for a variable storage time 
(typically 25 to 60 ms). This time is long enough for untrapped atoms to leave the detection region. 
The primary MOT light 
and the repumper are switched on again 
for a period of approximately 1 ms to detect the 
trapped atoms via spatially resolved detection of fluorescence with a spatial
resolution of 17 $\mu$m (rms-spread of smallest observed structures). 

We obtain a two-dimensional array of approximately 
80 well separated dipole traps with a potential depth of about 1 mK 
containing up to $10^3$ atoms 
(Fig. \ref{ARRAYFIG1} (a)). 
The number of filled traps is limited by the size of the laser beam illuminating
the microlens array and by the initial MOT size.
The apparent larger extent of the individual traps in the horizontal direction in all
images presented in this paper is caused by the detection optics 
being horizontally tilted relative to the beam axis of the trap light,
necessary to avoid trapping light entering the camera aperture. 
The detection efficiency of our setup is already high enough to be able to detect atom 
samples of fewer than 100 atoms per trap. We are currently optimizing the detection 
efficiency to allow the observation of single atoms as well \cite{SCHL01,singleatom,KUHR01}.
 
Illumination of only one row of the microlens array 
leads to a one-dimensional array of dipole traps (Fig. \ref{ARRAYFIG1} (b)).  
For the traps of this array (power per trap P = 3 mW, $\Delta\lambda$ = 0.4 nm) 
the calculated potential depth is $U_0/k_B$ = 2.5 mK, which agrees within a factor of 2 
with the one inferred from the measured radial 
oscillation frequency of 7.5 kHz \cite{Footnote1}. The discrepancy can be fully 
explained by the known uncertainties in determining the laser power per trap, the
focal waist, and the oscillation frequencies.
The lifetime of the atoms 
in the traps is 
35 ms, which is most probably limited by heating due to scattering of residual near resonant light 
not completely absorbed from the trapping beam. 
Using a time-of-flight technique, we determined the atom temperature to be
below 20 $\mu$K, 
which suggests the presence of an additional cooling mechanism during the loading phase as also 
observed in
 \cite{CHAP01}.


%

\begin{figure}
   \begin{center}
   \parbox{6cm}{
   \epsfxsize 6cm
   \epsfbox{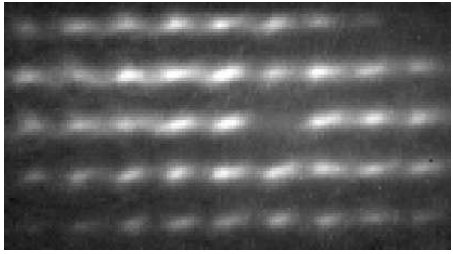}
}
   \end{center}
   
   \caption{Demonstration of the selective addressability of individual trap sites: By 
   focusing a near-resonant laser beam onto one of the dipole traps (row 3, column 6) 
   during the storage period, 
   the atoms in this trap
   are removed, while the other dipole traps remain unaffected.}        
    \label{ARRAYFIG3}
\end{figure}

In addition to its scalability, our approach is especially suited to fulfill 
another requirement for the physical implementation of quantum 
information processing, namely the ability to selectively address, initialize, and read out 
individual qubits: The large lateral separation between the dipole traps 
enables us to selectively address the individual traps in a straightforward 
fashion.
We demonstrate this by focusing a near-resonant 
laser beam onto one of the dipole traps for a few ms after the 
loading process is 
completed \cite{WEITZ}. This 
heats the atoms out of the addressed dipole trap.
As can 
be seen in Figure \ref{ARRAYFIG3}, no atoms are left at the site of the addressed dipole 
trap, while the atoms at the adjacent sites remain unaffected.
By two-dimensional scanning of the addressing beam 
or by illuminating each lenslet individually with spatially modulated addressing light, 
every site can be addressed individually.
This opens the possibility to selectively prepare and manipulate the qubits in the 
individual 
traps. 

As a next step, we demonstrated the site specific and state selective initialization and readout of 
atomic 

\begin{figure}
   \begin{center}
   \parbox{6.5cm}{
   \epsfxsize 6.5cm
   \epsfbox{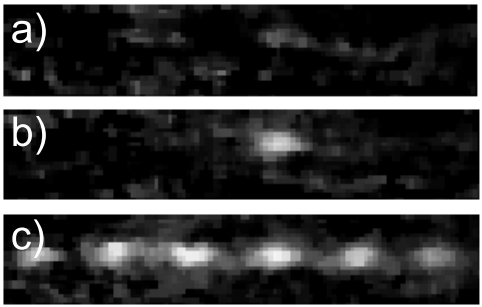}     
}
   \end{center}
   
   \caption{Site specific and state selective initialization 
   and read out of quantum states. 
   All sites of a one-dimensional array 
   of dipole traps are filled with atoms. For detection only light resonant with the 
   $5S_{1/2} (F=3) \rightarrow 5P_{3/2} (F'=4)$ transition is applied. (a) All atoms 
   are in the F=2 state, they do not scatter the detection light. (b) The atoms at one site 
   have been transfered to the F=3 state and can be seen. (c) The atoms at all sites 
   have been transfered to the F=3 state.}        
    \label{ARRAYFIG4}
\end{figure}

quantum states (Fig. \ref{ARRAYFIG4}).
Here, we illuminate a one-dimensional atom array (analogous to Fig. \ref{ARRAYFIG1} (b)) 
only with 
light resonant with the $5S_{1/2} (F=3) \rightarrow 5P_{3/2} (F'=4)$ transition 
(i.e. the repump light switched off) 
during detection. Since the atoms are almost 
exclusively in the 
lower hyperfine groundstate $5S_{1/2} (F = 2)$ after the loading 
phase, they do not scatter 
the detection light (Fig. \ref{ARRAYFIG4} (a)) unless we actively pump them into the 
upper hyperfine groundstate $5S_{1/2} (F = 3)$ during the time the atoms are stored in the 
dipole traps, i.e. prior to the detection phase (Fig. \ref{ARRAYFIG4} (b),(c)). This has been done 
for one 
(Fig. \ref{ARRAYFIG4} (b)) or, alternatively, all (Fig. \ref{ARRAYFIG4} (c)) of the trap sites. 
This demonstrates the site specific and state selective initialization and detection 
capability of our approach. 
 
\begin{figure}
   \begin{center}
   \parbox{8cm}{
   \epsfxsize 8cm
   \epsfbox{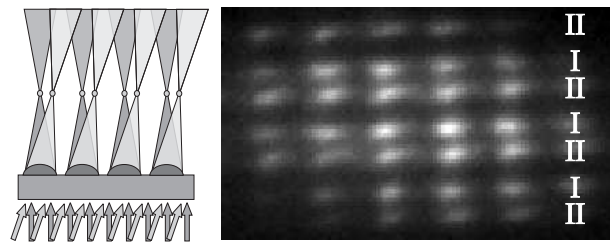}
}
   \end{center}
   
   \caption{Two interleaved sets of arrays of trapped atoms (Arrays I and II at right) 
   obtained by illuminating one microlens array with 
   two beams under slightly different angles (left). The distance between traps 
   of the two arrays is approximately 45 $\mu$m.}        
    \label{ARRAYFIG5}
\end{figure}

\begin{figure}
   \begin{center}
   \parbox{4cm}{
   \epsfxsize 4cm
   \epsfbox{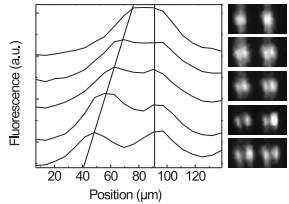}
}
   \end{center}   
   \caption{(left) Cross-sections through pairs of dipole traps as in 
   Fig. \ref{ARRAYFIG5} (rotated by 90$^{\circ}$)
   for different separations between the traps 
   ranging from 18 $\mu$m to 45 $\mu$m. 
   (right) Image sections  
   showing two pairs of traps with decreasing separation
   from bottom to top corresponding to the cross-sections at left.}        
    \label{ARRAYFIG6}
\end{figure}

While many of the advantages of our system
result from the large lateral 
separation of the individual trap sites, it is also possible to actively control the 
distance between individual traps if smaller or adjustable distances are 
required, e.g.
for quantum gate operations and the entanglement of atoms via atom-atom interactions. 
In our setup, 
this can be accomplished by illuminating one microlens array with two 
beams under slightly different angles (Fig. \ref{ARRAYFIG5} (left))  \cite{BIRK01,SCHL01}, 
which results in two interleaved sets 
of arrays of trapped atoms. 
Interference effects between the two laser beams are prevented by using orthogonal linear polariations.
Figure \ref{ARRAYFIG5} (right) shows two vertically displaced sets of arrays of trapped atoms 
with a mutual separation of 45 $\mu$m.
The separation only depends 
on the angle between the two laser beams and can easily be changed, especially to smaller 
values.
By reducing the relative angle to zero, overlapping traps can be 
created.
Figure \ref{ARRAYFIG6} shows vertical cross-sections through one pair of dipole 
traps for different angles between the two laser beams and thus different site separations. 
We could also 
change the separation of the traps and thus move the atoms in one set of traps in real time by deflecting   
one beam with the help of a fast acoustooptical beam deflector.
With this technique it
should become possible to selectively prepare qubits at large 
distances, then 
decrease the distance for gate operations and entanglement 
via atom-atom interactions and then increase the 
distance again for the read out of the qubit states. 

In order to demonstrate the full potential for quantum information
processing, we evaluate the criteria given in \cite{DIVI00} for our approach.
Scalability, site specific and state selective initialization and readout,
as well as the ability to change the separation of trapping sites at will are the most important
advantages of our approach, and have been demonstrated in the previous sections.
In addition, the concept presented here and the underlying technology fulfill
the remaining two criteria, namely long relevant decoherence times and the availabilty
of a universal set of quantum gates as well. 

With the specific lens array of this work (beam waist 7 $\mu$m) and 
a standard set of laser parameters (Ti:Sa laser, wavelength 825 nm, 
power per trap 50 mW) several tens of traps with radial vibrational frequencies 
of 10 kHz and a decoherence
time (assumed to be limited by spontaneous scattering) of 50 ms 
can be obtained \cite{BIRK01}. 
These parameters allow the implementation of collisional
gates \cite{JAKS99} with a decoherence time being at least 50 times longer 
than the gate time. 

The full potential of our concept can be exploited by using 
available lens arrays optimized for large numerical 

aperture: 
it has been shown experimentally  \cite{HESS97}
that beam waists below 1 $\mu$m can be achieved with microlens arrays
similar to the one used in this work. 
With standard laser parameters 
(wavelength 850 nm, power per trap 1 mW) radial vibrational frequencies of 
50 kHz and decoherence times of 150 ms can be obtained.
This 
allows the implementation of various types of quantum gates \cite{JAKS99,JAKS00,CHAR02}. 
Due to
its potentially short gate times and its insensitivity to the temperature of the
atoms and to the variations in atom-atom separation, the 
Rydberg gate of \cite{JAKS00} deserves specific attention: 
Using the parameters and scaling laws 
of \cite{JAKS00}, for atoms in microlens dipole traps with a waist and a respective 
minimum trap separation of 
1 $\mu$m, a gate time of about 1 $\mu$s can be achieved. 
This gate time is 20 times shorter
than the oscillation period and $10^4$ times shorter than 
the decoherence time, giving favorable experimental conditions. 
Further evidence for the potential of this approach can be drawn from the fact that
this array of 1 $\mu$m-waist dipole traps represents a 2D extension of the dipole trap
demonstrated in \cite{SCHL01}. For this trap a detailed investigation \cite{PROT02}
gives proof for the feasability of Rydberg gates and explicitly shows 
that for trap separations of 1 to 5 $\mu$m, gate times of 1 to 10 $\mu$s are possible.

We conclude this discussion by pointing out that since our dipole trap arrays
can give the same trap parameters as single dipole traps or standing wave dipole potentials,
the single atom loading schemes experimentally demonstrated in \cite{SCHL01} and \cite{KUHR01}
can be extended to our configurations in a straightforward fashion. 
In addition, with the achievable vibrational frequencies being larger than the recoil
frequency, sideband-cooling of atoms to the vibrational ground state of 
the trap arrays should be possible.
Finally, the demonstration of the loading of
periodic dipole potentials with single atoms from a
Bose Einstein condensate \cite{MOTT02} might allow an extemely efficient
means of single atom loading into vibrational ground states of our trap arrays.

This work is supported by the SFB 407 and the program {\it 
Quanteninformationsverarbeitung} 
of the {\it Deutsche Forschungsgemeinschaft} as well as by the program ACQUIRE 
(IST-1999-11055) of the {\it European Commission}. 



\begin{references}
\bibitem{Footnote1} We measure this frequency by parametrically heating the atoms
 and observing the number of remaining atoms.
\end{references}
\end{document}